\documentclass[a4paper,11pt]{article}
\usepackage{pos}

\newcommand{\gcn}[1]{\href{https://gcn.gsfc.nasa.gov/gcn3/#1.gcn3}{GCN {#1}}}
\newcommand{\atel}[1]{\href{http://www.astronomerstelegram.org/?read=#1}{ATel {#1}}}

\title{Realtime Follow-up of Astrophysical Transients with the IceCube Neutrino Observatory}
 \ShortTitle{IceCube Fast Response}

\author{The IceCube Collaboration \\{\normalsize \normalfont(a complete list of authors can be found at the end of the proceedings)}}

\emailAdd{apizzuto@icecube.wisc.edu}

\abstract{Realtime analyses are necessary to identify the source of high energy neutrinos. As an observatory with a 4$\pi$ steradian field of view and near-100\% duty cycle, the IceCube Neutrino Observatory is a unique facility for investigating transients. In 2016, IceCube established a pipeline that uses low-latency data to rapidly respond to astrophysical events that were of interest to the multi-messenger observational community. Here, we describe this pipeline and summarize the results from all of the analyses performed since 2016. We focus not only on those analyses which were performed in response to transients identified using other messengers such as photons and gravitational waves, but also on how this pipeline can be used to constrain populations of astrophysical neutrino transients by following up high-energy neutrino alerts.

\vspace{4mm}
{\bfseries Corresponding authors:}
Alex Pizzuto$^{1*}$, Abhishek Desai$^{1}$, and Raamis Hussain$^{1}$\\
{$^{1}$ \itshape University of Wisconsin-Madison and Wisconsin IceCube Particle Astrophysics Center}\\[4mm]
$^*$ Presenter

\FullConference{37$^{\rm{th}}$ International Cosmic Ray Conference (ICRC 2021)\\
		July 12th -- 23rd, 2021\\
		Online -- Berlin, Germany}
		}

\begin{document}
\maketitle

\section{Introduction}
The last several years have witnessed the first few associations between high-energy neutrinos and potential astrophysical counterparts. Many of these associations -- such as that between the flaring blazar TXS 0506+056 and a high-energy neutrino alert event IceCube-170922A \cite{IceCube:2018dnn,IceCube:2018cha}, as well as high energy photons with gravitational waves (GW170817 / GRB170817A) \cite{2017ApJ...848L..12A}  --  were made possible by advancement in low-latency analysis pipelines. While many such tools rely on the use of high-energy neutrino alerts \cite{Blaufuss:2019fgv} to trigger multi-wavelength followups, realtime neutrino astronomy does not necessitate solely the use of ``alert'' quality events. As IceCube simultaneously observes the entire sky with over 99\% duty cycle, it is a unique facility poised not only to issue its own alerts but also to respond to interesting astrophysical phenomena.

IceCube is a cubic-kilometer neutrino detector \cite{Aartsen:2016nxy} instrumented at the geographic South Pole. It does not observe neutrinos directly, but instead uses an array of 5160 digital optical modules (DOMs) buried in the ice to detect the Cherenkov light emitted by the secondary charged particles produced in neutrino-nucleon interactions. The direction and energy of the neutrino can then be estimated by reconstructing the secondary particles from the light collected by the DOMs. While sensitive to neutrinos from cosmic sources, the majority of the events detected by IceCube come from atmospheric backgrounds. 

\begin{figure}
    \centering
    \includegraphics[width=0.98\textwidth]{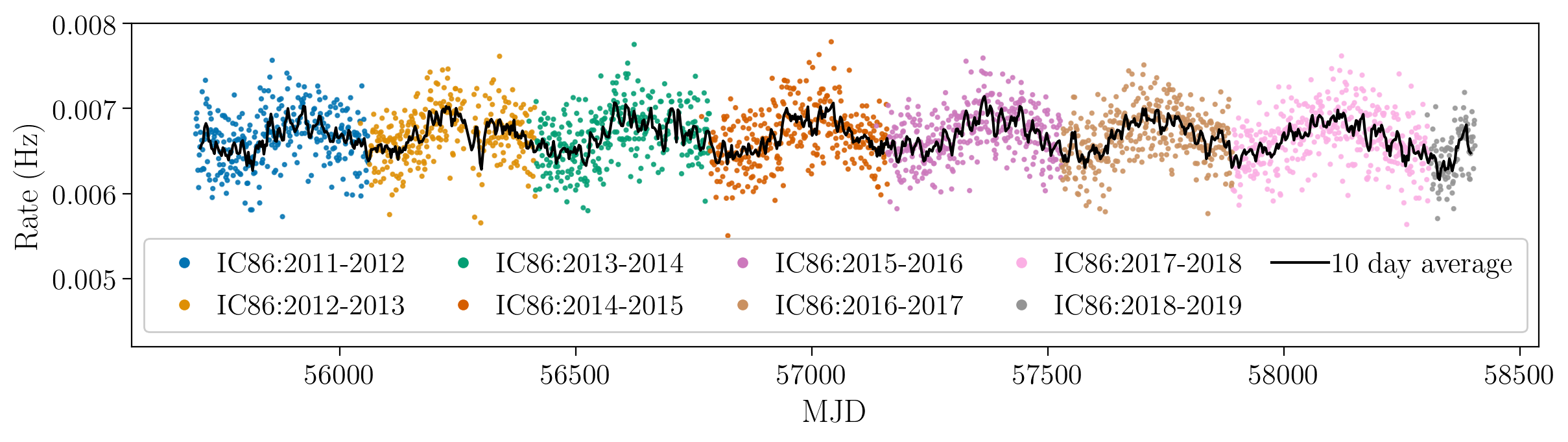}
    \caption{All-sky event rate of the event selection used for this analysis as a function of time. Different detector operation seasons are denoted by different colors, where ``IC86'' denotes the full 86 string detector configuration for IceCube. Each data point is the rate calculated from averaging 3 sequential 8-hour ``runs.'' In addition to a purely statistical fluctuation on the order of 5\%, the overall rate displays a clear annual modulation, whose peak to peak amplitude is approximately 4\% of the mean rate.}
    \label{fig:gfu_rate}
\end{figure}

In order to expedite the identification of cosmic neutrino sources, whenever IceCube detects a neutrino event with a high probability of astrophysical origin, it issues an alert to the observational community with low-latency. This alert infrastructure was established in 2016 \cite{Aartsen:2016lmt} and was later improved in 2019 \cite{Blaufuss:2019fgv}, and now sends an average of $\sim$10 ($\sim$20) events per year of high (moderate) probability of astrophysical origin, and is most sensitive to astrophysical neutrinos with energies of a few hundred TeV. In addition to these few events per year, there is another stream of data that is sent from the South Pole to computing centers at the University of Wisconsin-Madison with low-latency comprised of muon-neutrino candidate events. This data stream is called the ``Gamma-ray FollowUp'' (GFU) selection for its initial use of triggering rapid very-high-energy gamma-ray observations \cite{Kintscher:2016uqh}, and it has an all-sky rate of $\sim$6 mHz. Though this rate is dominated by atmospheric backgrounds and is much larger than that of alert event rates, it also comes with the benefit of a much larger effective area. It also is sensitive to much lower energies than the alert event sample, for example, for a source in the northern sky with an intrinsic spectrum of $dN/dE \propto E^{-2}$, 90\% of the true energies of detected neutrinos from this source would fall between roughly 1 TeV and 500 TeV. The background rate of the GFU sample has modest annual modulation due to seasonal variation of the atmospheric temperature and pressure, which can be seen in Figure~\ref{fig:gfu_rate}. 

Here, we discuss the IceCube ``Fast Response Analysis,'' which uses the GFU sample to search for neutrino emission from astrophysical transients in real time. This pipeline has been in use since 2016, and has been used to both respond to interesting transients -- such as bright gamma-ray bursts (GRBs), fast radio bursts (FRBs), and intense blazar flares -- identified with messengers besides neutrinos as well as to look for lower energy neutrino emission in the direction of alert events. This pipeline has also been used to search for neutrinos from gravitational wave progenitors, though now an independent pipeline exists to use IceCube data to respond to gravitational wave triggers \cite{Veske:2021icrc}. In section~\ref{sec:analysis}, we describe the analysis pipeline, then we summarize our current results for using the pipeline to search for neutrino emission from bright transients in section~\ref{sec:external}. In section~\ref{sec:internal} we outline the science case for using this pipeline to search for neutrino emission in the direction of neutrino alerts. For more analysis details and for a more thorough discussion of the interpretation of analyses performed in response to external triggers, see \cite{Abbasi:2020aae}.

\section{Analysis Method}
\label{sec:analysis}
The Fast Response Analysis (hereafter FRA) uses the same unbinned maximum likelihood analysis that is a cornerstone in many searches for transient sources of astrophysical neutrinos, described in full in \cite{Abbasi:2020aae}. As this analysis is used to search for short timescale neutrino emission, the likelihood form used is that of an ``extended likelihood'' which features a Poisson term that compares the total number of events observed in the search time window with the number expected from background. This has also been used in searches for neutrinos from e.g. GRBs \cite{Aartsen:2017wea} and FRBs \cite{Aartsen:2019wbt}. The likelihood also includes an energy term, optimized for an assumed signal spectral shape of $dN/dE \propto E^{-2}$, in order to weight higher energy events as more signal-like compared to lower energy events which are more consistent with atmospheric background. Although the analysis is optimized for an $E^{-2}$, the analysis remains sensitive over a broad range of power law indices, and has been explicitly tested with injected assumed spectra of $E^{-2.5}$ and $E^{-3}$.

\begin{figure}
    \centering
    \includegraphics[width=0.98\textwidth]{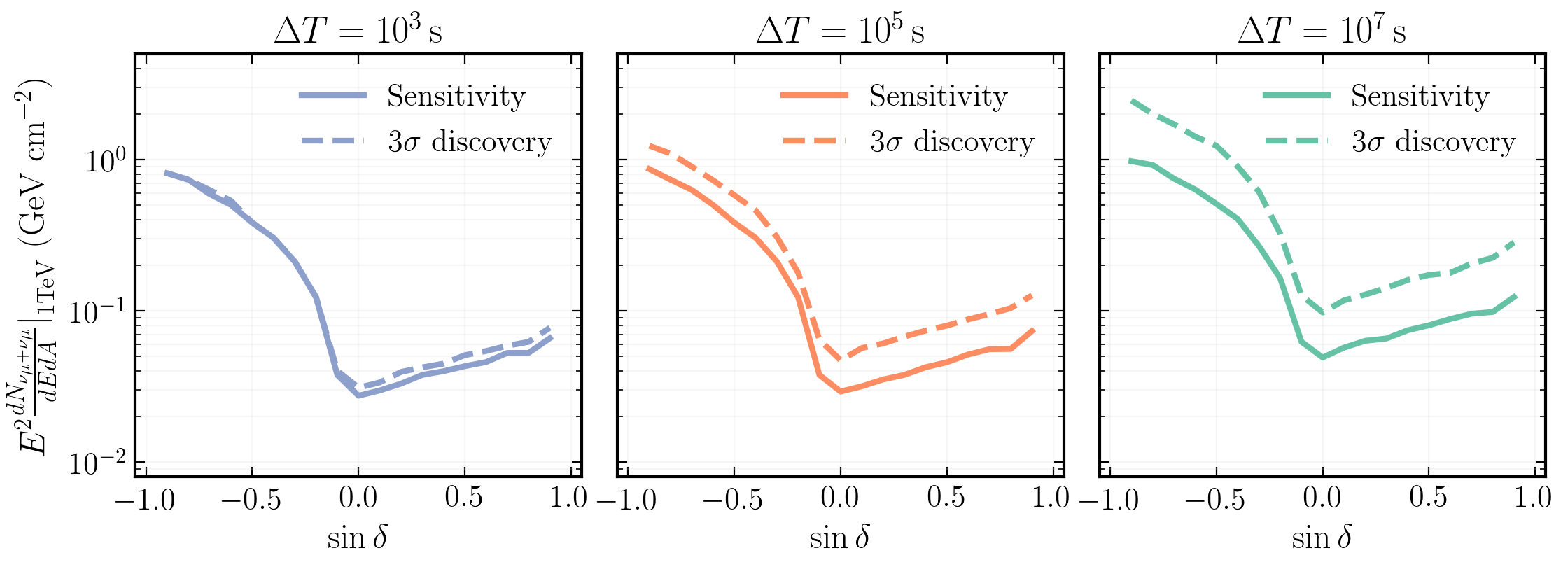}
    \caption{Analysis sensitivity as a function of declination ($\delta$) for three different time scales, under the assumption of an $E^{-2}$ power-law spectrum. The number of coincident neutrino candidate events from atmospheric background increases as the time window for the analysis increases, which in turn degrades the sensitivity and discovery potential.}
    \label{fig:sensitivity}
\end{figure}

The analysis sensitivity -- defined as the median 90\% upper limit that would be set under the assumption of the null hypothesis -- for three different analysis time windows is displayed in Figure~\ref{fig:sensitivity}. These analysis time windows were selected to show the range of typical timescales that we consider when performing an analysis. The analysis is most sensitive near the Equator and in the Northern Celestial hemisphere, where the Earth attenuates the atmospheric muon background. However, the analysis is still sensitive to bright transients in the Southern Hemisphere. With an average background rate of $\sim$6~mHz and an average uncertainty on the direction of reconstructed neutrino candidate events of $\sim$1$^{\circ}$, for time-windows less than $10^5$~s, we do not expect many background events to be spatially coincident with a the source of interest for the analysis. As such, the sensitivity is constant for these short time windows. In this regime, the analysis is capable of yielding a result that is significant at the 3$\sigma$ level, pre-trials, from just one coincident signal event. For larger time windows, the sensitivity degrades, but the analysis has been used to search for neutrino emission on timescales as long as one month. 

The analysis is most sensitive when searching for neutrino emission from a well-localized point source. However, for sources with localization uncertainty, such as GRBs identified with the \textit{Fermi}-GBM or for searching for low energy neutrino emission in the direction of poorly localized neutrino alert events, the uncertainty can be taken into account by searching for neutrino point sources on the whole sky and combining this point source information with information from the localization PDF of the source being investigated. This is the same method as is used when searching for neutrino counterparts to gravitational wave events \cite{Aartsen:2020mla}.

\section{External Triggers}
\label{sec:external}
The initial purpose of the FRA was to respond to interesting astrophysical transients that were promising sources of neutrinos. Since 2016, the FRA has been used to search for neutrino emission from sources such as GRBs, FRBs, localized GW progenitors, bright blazar flares, and fast blue optical transients. So far, 62 analyses have been performed in response to external triggers, the first 58 of which are discussed in \cite{Abbasi:2020aae}. The most recent 4 analyses have all resulted in non-detections, all resulting in a $p$-value of 1.0. The most recent analyses are summarized in Table~\ref{tab:new_res}. Three of these analyses were searching for neutrino emission from bright GRBs, all of which used an analysis time window of one hour, beginning 10 minutes prior to the GRB trigger. The other analysis was searching for neutrino emission from SGR~1935+2154, which was previously associated an FRB \cite{2020arXiv200510324T,2020arXiv200510828B}. In October 2020, CHIME announced the detection of a flurry of FRBs from the direction of SGR 1935+2154 by CHIME (\atel{14074}), and shortly thereafter Swift/BAT announced a possible coincident x-ray flare (\atel{14075}). The FRA was used to search for neutrino emission in the case that these were associated, and looked for neutrinos with a time window one day in duration, centered on the first FRB trigger. No neutrino emission was detected, and the x-ray observations were later shown to be due to a temporary detector glitch (\atel{14076}), but the FRA is well-poised to search for neutrino emission from future FRBs.

\begin{table}[h]
    \centering
    \begin{tabular}{l|c|c|c}
         Source Name & Start Date & Analysis $p$-value & Upper limit ($\times 10^{-2}$ GeV cm$^{-2}$)\\ 
         \hline \hline
         GRB 200729A & 2020-07-29 & 1.0 & 5.3 \\
         SGR 1935+2154 & 2020-10-07 & 1.0 & 4.2 \\
         GRB 201015A & 2020-10-15 & 1.0 & 5.9 \\
         GRB 201216C & 2020-12-16 & 1.0 & 4.0 
    \end{tabular}
    \caption{Summary of externally triggered FRA results since July 2020. To date, no significant detections have been made. Upper limits are placed on the energy-scaled time-integrated neutrino flux, $E^2 \Delta T dN/dE$, under the assumption of an $E^{-2}$ power law. IceCube publicly issued the results of the searches for neutrinos from GRB~200729A (\gcn{28173}) and GRB~201216C (\atel{14277}).}
    \label{tab:new_res}
\end{table}

\section{Internal Triggers}
\label{sec:internal}
In addition to searching for neutrinos from identified astrophysical transients, the FRA can also be used to search for additional neutrino events that are spatially coincident with IceCube alert events. When following up an alert event, the uncertainty on the alert event localization can be incorporated using the technique outlined in Section~\ref{sec:analysis}, and the alert event is excluded from the analysis. This technique is especially useful as the GFU selection has a much larger effective area than that of the alert event sample, especially for low energies. For example, consider the ratio of the expected number of signal events in the GFU sample compared to the expected number of alert events,

\begin{equation}
\label{eq:ev_ratio}
    \frac{\langle N_{\mathrm{GFU}} \rangle}{\langle N_{\mathrm{alert}} \rangle} = \frac{\int \phi(E) A^{\mathrm{eff}}_{\mathrm{GFU}}(\delta, E) dE }{\int \phi(E) A^{\mathrm{eff}}_{\mathrm{alert}}(\delta, E) dE}\; ,
\end{equation}
for a source at declination $\delta$ with a spectrum $\phi = \phi_0 E^{-\gamma}$, where $A^{\mathrm{eff}}$ is the energy and declination dependent effective area. For a source at $\delta=0^{\circ}$ with $\gamma=2$ ($\gamma=2.5$), this ratio is 13 (97). For a source in the northern sky with $\delta=30^{\circ}$, this ratio is 27 (40), meaning depending on the location on the sky and intrinsic source spectrum, we can expect tens of astrophysical neutrino events in the GFU sample for every expected alert event.

Current limits on neutrino point sources indicate that for most neutrino sources $\langle N_{\mathrm{alert}} \rangle < 1$ \cite{Strotjohann:2018ufz}, meaning that when a neutrino alert is coincident with a source, there is significant uncertainty introduced when trying to measure the true flux of the source. By searching for additional neutrino events in the GFU sample from the direction of these alert events, we can leverage the larger effective area of the GFU sample to more accurately measure fluxes from these objects. While the FRA has traditionally been used to search for sources in real time (and can be used to follow up future alerts in real time), using this analysis on an archival sample of alert events could help inform us about the population properties of astrophysical neutrino sources, including their number density and luminosity function.

To test how sensitive the FRA is for searching for populations of astrophysical neutrino sources, we simulated populations of neutrino sources using the publicly available code FIRESONG (FIRst Extragalactic Simulation Of Neutrinos and Gamma rays) \cite{Tung2021}. FIRESONG can, given a number density and average luminosity for a population of neutrino sources, simulate the full population and calculate the neutrino flux from each source, taking into account factors such as cosmic evolution and luminosity functions. For each source, we calculate $\langle N_{\mathrm{alert}} \rangle$, and sample from a Poisson distribution to see which sources yield alert events. For those sources, we also calculate $\langle N_{\mathrm{GFU}} \rangle$ to find the number of additional events to inject into the analysis. Alert events arising from atmospheric backgrounds are also simulated with rates according to \cite{Blaufuss:2019fgv}. 

\begin{figure}
    \centering
    \includegraphics[width=0.49\textwidth]{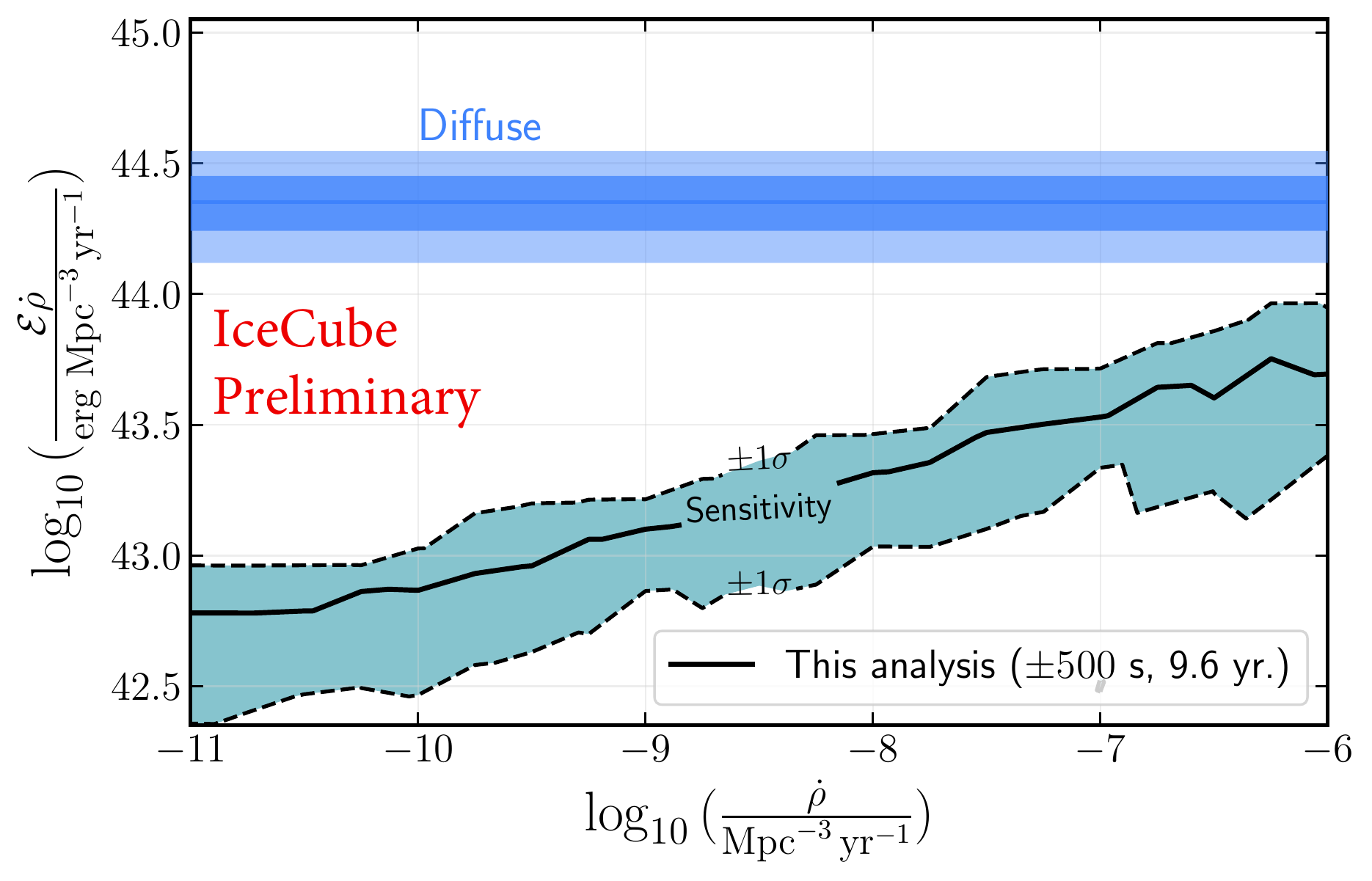}
    \includegraphics[width=0.49\textwidth]{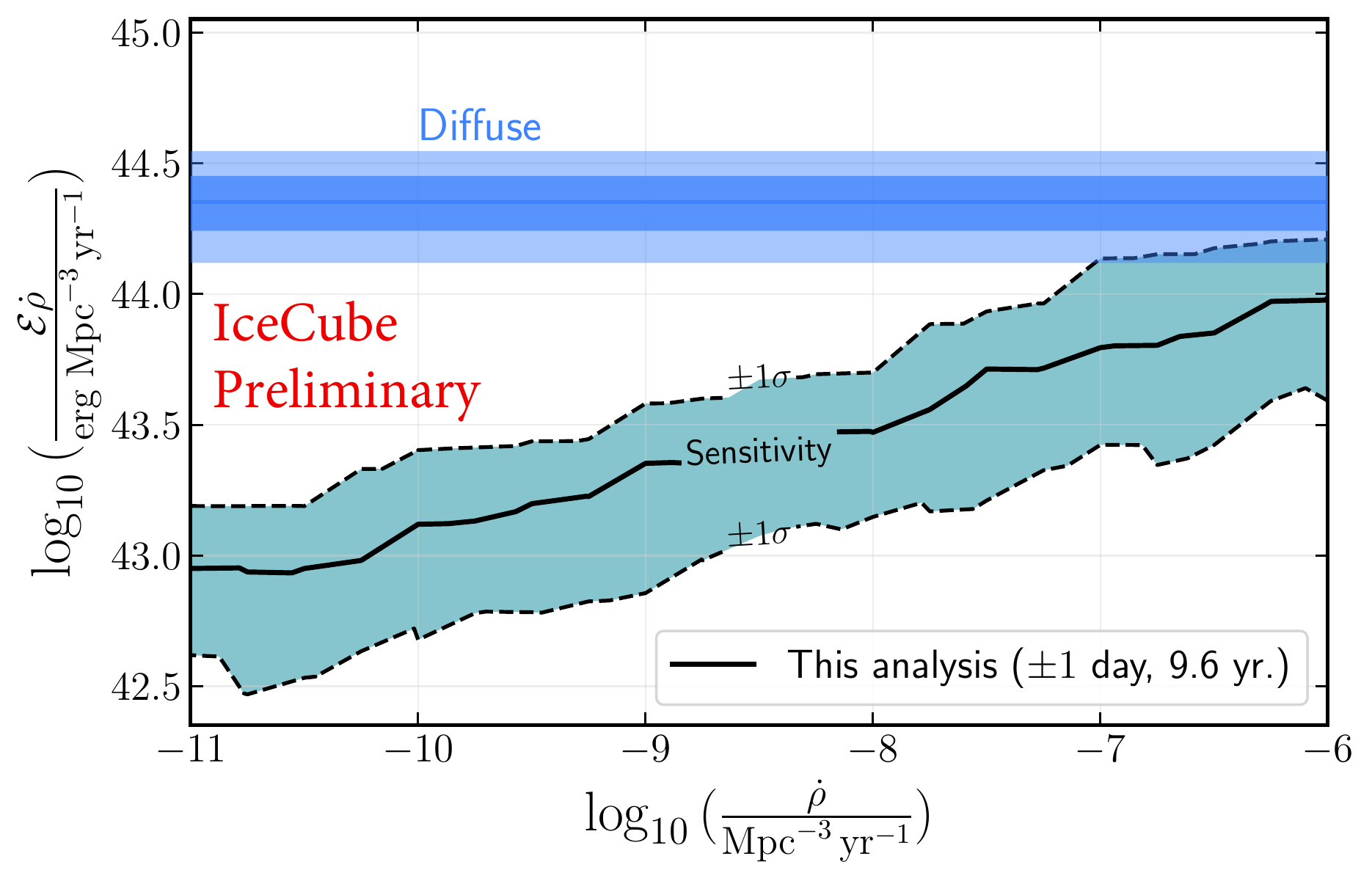}
    \caption{Sensitivity to populations of transients when using the FRA to search for additional neutrino events in the direction of IceCube alert events. The analysis sensitivity is shown for two different timescales, $\pm 500$ s (left) and $\pm 1$ day (right). A population that would saturate the diffuse astrophysical neutrino flux is shown in the blue band. The sensitivity of this analysis (median upper limit in the case of a non-detection) is shown in black, and the green band shows how much the upper-limit can be expected to fluctuate on average in the case of a non-detection. For this calculation, we assume there are 9.6 years worth of alert events which can be followed up.}
    \label{fig:pop_sens_transient}
\end{figure}

The sensitivity of using the FRA to search for neutrino emission in the direction of alert events is shown in Figure~\ref{fig:pop_sens_transient} in terms of the average isotropic energy emitted by each source between 10 TeV and 10 PeV, $\mathcal{E}$, assuming sources have spectra consistent with the measured diffuse flux, $dN/dE \propto E^{-2.5}$, and assuming 9.6 years of online operations. While we use an assumed spectrum of $E^{-2}$ when searching for neutrinos from external triggers, we fix the index in our likelihood to $E^{-2.5}$ for internally triggered analyses to focus on sources that could explain contributions to the diffuse astrophysical neutrino flux. For these sensitivities, we inject transient neutrino source populations with a standard candle luminosity function and assume a redshift evolution that tracks star formation consistent with the rate derived in \cite{Madau:2014bja}. We show the sensitivity for two different timescales, $\pm 500$ s around the alert time and $\pm 1$ day around the alert time. These time windows were chosen to strike a balance between model independence and background control: while larger search windows are sensitive to signal hypotheses up to the search window duration, they come at the cost of increased background. 

When compared to the flux which would saturate the total measured diffuse astrophysical neutrino flux, the analysis is most sensitive to rare populations of sources, where sources, though less numerous, are on average brighter. For the analysis time window of $\pm 500$ s ($\pm 1$ day), the analysis is sensitive to populations of sources with a rate density of $\dot{\rho} = 10^{-11}$~Mpc$^{-3}$~yr$^{-1}$ at the level of 3\% (4\%) of the diffuse flux. For a more numerous source population with a rate density of $\dot{\rho} = 10^{-6}$~Mpc$^{-3}$~yr$^{-1}$, the analysis is sensitive to a population producing 30\% (45\%) of the diffuse flux.

Although the FRA has been used in the past for short timescale transients, the methodology outlined above for searching for populations of transient neutrino sources can also be used to search for neutrino sources which are constantly emitting as a function of time. The sensitivity of such an analysis is comparable to the short timescale cases, but the increased background that comes from analyzing larger time windows degrades the sensitivity. As the hypothesis tested here is one of constant emission, we instead show the sensitivity in terms of the luminosity, $\mathcal{L}$, (instead of emitted energy for the transient case) of each source between 10 TeV and 10 PeV, as a function of the local density of neutrino sources, $\rho$. In both the short timescale and time-integrated cases, the analysis is sensitive to rare populations of sources that significantly contribute to the diffuse astrophysical neutrino flux.

\begin{figure}
    \centering
    \includegraphics[width=0.65\textwidth]{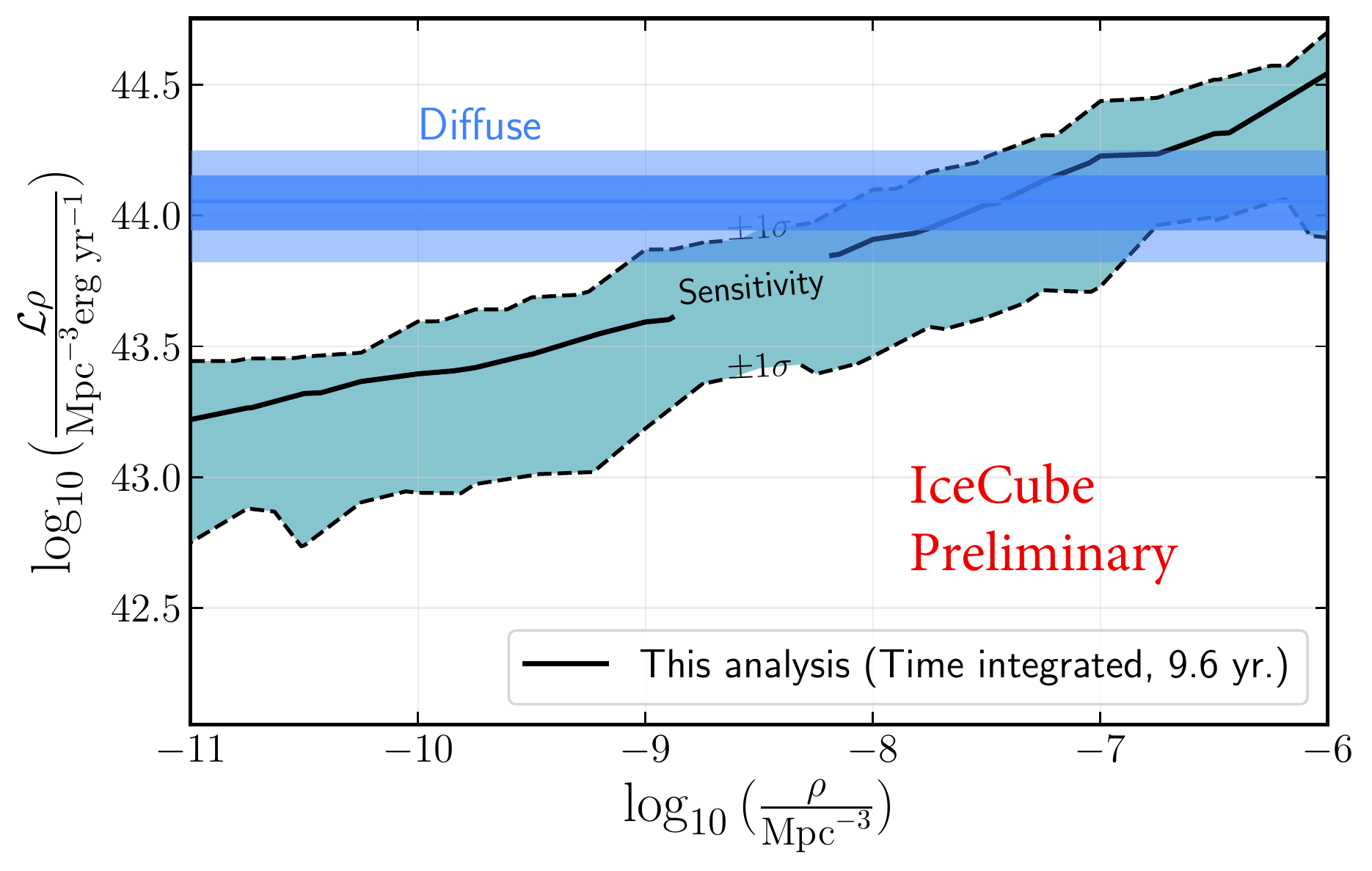}
    \caption{Sensitivity to populations of steady neutrino sources when searching for neutrino emission in the direction of IceCube alert events. The colors are the same as in Figure~\ref{fig:pop_sens_transient}.}
    \label{fig:pop_sens_steady}
\end{figure}

As such, we plan to use this analysis in two different manners. First, we will apply the methods outlined above to all archival events that pass current alert quality cuts \cite{Blaufuss:2019fgv}, but include events that predate the operation of the realtime alert stream. This sample of alert events, which will be the focus of a future work, contains a total of 275 alert events from 9.6 years of operation between calendar years 2011 and 2020. Second, to help identify astrophysical neutrino sources in real time, we will apply the methods outlined above to follow up future alert events in real time. This will help identify which future alert events are coming from bright neutrino sources.

\section{Conclusion}

Thanks to IceCube's $4\pi$ steradian field of view and near 100\% duty cycle, it is an observatory well equipped to constrain astrophysical transients. The low-latency pipeline capabilities of IceCube make it possible to constrain such sources in real time. Here, we described the Fast Response Analysis, which is a tool used to search for neutrino transients in real time. The analysis has been used in the past to search for neutrinos from transients identified with other messengers, and though no significant detections have been made, it has been used to set limits on bright transients that have long been thought to be promising neutrino emitters. In addition to responding to external alerts, the Fast Response Analysis is also able to search for low energy neutrino emission spatially coincident with IceCube alert events. The results from such an analysis would be informative for discerning the statistical characteristics of neutrino source populations which contribute significantly to the diffuse astrophysical flux. In the future, this analysis could identify in real time those neutrino alerts which are accompanied by lower energy astrophysical neutrinos, which may represent the brightest subpopulation of astrophysical neutrino sources.

\bibliography{references}

\providecommand{\href}[2]{#2}\begingroup\raggedright\begin{thebibliography}{10}

\bibitem{IceCube:2018dnn}
{\bfseries IceCube, Fermi-LAT, MAGIC, AGILE, ASAS-SN, HAWC, H.E.S.S., INTEGRAL,
  Kanata, Kiso, Kapteyn, Liverpool Telescope, Subaru, Swift NuSTAR, VERITAS,
  VLA/17B-403} Collaboration, M.~G. Aartsen {\em et~al.}
  \href{http://dx.doi.org/10.1126/science.aat1378}{{\em Science} {\bfseries
  361} no.~6398, (2018) eaat1378}.

\bibitem{IceCube:2018cha}
{\bfseries IceCube} Collaboration, M.~G. Aartsen {\em et~al.}
  \href{http://dx.doi.org/10.1126/science.aat2890}{{\em Science} {\bfseries
  361} no.~6398, (2018) 147--151}.

\bibitem{2017ApJ...848L..12A}
B.~P. {Abbott} {\em et~al.}
  \href{http://dx.doi.org/10.3847/2041-8213/aa91c9}{{\em The Astrophysical
  Journal Letters} {\bfseries 848} (Oct., 2017) L12}.

\bibitem{Blaufuss:2019fgv}
{\bfseries IceCube} Collaboration, E.~Blaufuss, T.~Kintscher, L.~Lu, and C.~F.
  Tung \href{http://dx.doi.org/10.22323/1.358.1021}{{\em PoS} {\bfseries
  ICRC2019} (2020) 1021}.

\bibitem{Aartsen:2016nxy}
{\bfseries IceCube} Collaboration, M.~G. Aartsen {\em et~al.}
  \href{http://dx.doi.org/10.1088/1748-0221/12/03/P03012}{{\em JINST}
  {\bfseries 12} no.~03, (2017) P03012}.

\bibitem{Aartsen:2016lmt}
{\bfseries IceCube} Collaboration, M.~G. Aartsen {\em et~al.}
  \href{http://dx.doi.org/10.1016/j.astropartphys.2017.05.002}{{\em Astropart.
  Phys.} {\bfseries 92} (2017) 30--41}.

\bibitem{Kintscher:2016uqh}
{\bfseries IceCube} Collaboration, T.~Kintscher
  \href{http://dx.doi.org/10.1088/1742-6596/718/6/062029}{{\em J. Phys. Conf.
  Ser.} {\bfseries 718} no.~6, (2016) 062029}.

\bibitem{Veske:2021icrc}
{\bfseries IceCube} Collaboration, D.~Veske {\em et~al.} {\em PoS} {\bfseries
  ICRC2021} (these proceedings) xyz.

\bibitem{Abbasi:2020aae}
{\bfseries IceCube} Collaboration, R.~Abbasi {\em et~al.}
  \href{http://dx.doi.org/10.3847/1538-4357/abe123}{{\em Astrophys. J.}
  {\bfseries 910} no.~1, (2021) 4}.

\bibitem{Aartsen:2017wea}
{\bfseries IceCube} Collaboration, M.~G. Aartsen {\em et~al.}
  \href{http://dx.doi.org/10.3847/1538-4357/aa7569}{{\em Astrophys. J.}
  {\bfseries 843} no.~2, (2017) 112}.

\bibitem{Aartsen:2019wbt}
{\bfseries IceCube} Collaboration, M.~G. Aartsen {\em et~al.}
  \href{http://dx.doi.org/10.3847/1538-4357/ab564b}{{\em Astrophys. J.}
  {\bfseries 890} no.~2, (2020) 111}.

\bibitem{Aartsen:2020mla}
{\bfseries IceCube} Collaboration, M.~G. Aartsen {\em et~al.}
  \href{http://dx.doi.org/10.3847/2041-8213/ab9d24}{{\em Astrophys. J. Lett.}
  {\bfseries 898} no.~1, (2020) L10}.

\bibitem{2020arXiv200510324T}
{\bfseries CHIME/FRB} Collaboration, B.~C. {Andersen} {\em et~al.}
  \href{http://dx.doi.org/10.1038/s41586-020-2863-y}{{\em Nature} {\bfseries
  587} no.~7832, (Nov., 2020) 54--58}.

\bibitem{2020arXiv200510828B}
C.~D. {Bochenek}, V.~{Ravi}, K.~V. {Belov}, G.~{Hallinan}, J.~{Kocz}, S.~R.
  {Kulkarni}, and D.~L. {McKenna}
  \href{http://dx.doi.org/10.1038/s41586-020-2872-x}{{\em Nature} {\bfseries
  587} no.~7832, (Nov., 2020) 59--62}.

\bibitem{Strotjohann:2018ufz}
N.~L. Strotjohann, M.~Kowalski, and A.~Franckowiak
  \href{http://dx.doi.org/10.1051/0004-6361/201834750}{{\em Astron. Astrophys.}
  {\bfseries 622} (2019) L9}.

\bibitem{Tung2021}
C.~F. Tung, T.~Glauch, M.~Larson, A.~Pizzuto, R.~Reimann, and I.~Taboada
  \href{http://dx.doi.org/10.21105/joss.03194}{{\em Journal of Open Source
  Software} {\bfseries 6} no.~61, (2021) 3194}.

\bibitem{Madau:2014bja}
P.~Madau and M.~Dickinson
  \href{http://dx.doi.org/10.1146/annurev-astro-081811-125615}{{\em Ann. Rev.
  Astron. Astrophys.} {\bfseries 52} (2014) 415--486}.

\end{thebibliography}\endgroup
\bibliographystyle{ICRC}

\clearpage
\section*{Full Author List: IceCube Collaboration}




\scriptsize
\noindent
R. Abbasi$^{17}$,
M. Ackermann$^{59}$,
J. Adams$^{18}$,
J. A. Aguilar$^{12}$,
M. Ahlers$^{22}$,
M. Ahrens$^{50}$,
C. Alispach$^{28}$,
A. A. Alves Jr.$^{31}$,
N. M. Amin$^{42}$,
R. An$^{14}$
% ...
\\

\noindent
% ...
$^{12}$ Universit{\'e} Libre de Bruxelles, Science Faculty CP230, B-1050 Brussels, Belgium \\
% ...
$^{14}$ Department of Physics and Laboratory for Particle Physics and Cosmology, Harvard University, Cambridge, MA 02138, USA \\
% ...
$^{17}$ Department of Physics, Loyola University Chicago, Chicago, IL 60660, USA \\
$^{18}$ Dept. of Physics and Astronomy, University of Canterbury, Private Bag 4800, Christchurch, New Zealand \\
% ...
$^{22}$ Niels Bohr Institute, University of Copenhagen, DK-2100 Copenhagen, Denmark \\
% ...
$^{28}$ D{\'e}partement de physique nucl{\'e}aire et corpusculaire, Universit{\'e} de Gen{\`e}ve, CH-1211 Gen{\`e}ve, Switzerland \\
% ...
$^{31}$ Karlsruhe Institute of Technology, Institute for Astroparticle Physics, D-76021 Karlsruhe, Germany  \\
% ...
$^{42}$ Bartol Research Institute and Dept. of Physics and Astronomy, University of Delaware, Newark, DE 19716, USA \\
% ...
$^{50}$ Oskar Klein Centre and Dept. of Physics, Stockholm University, SE-10691 Stockholm, Sweden \\
% ...
$^{59}$ DESY, D-15738 Zeuthen, Germany \\
% ...

\subsection*{Acknowledgements}

\noindent
USA {\textendash} U.S. National Science Foundation-Office of Polar Programs,
U.S. National Science Foundation-Physics Division,
U.S. National Science Foundation-EPSCoR,
Wisconsin Alumni Research Foundation,
Center for High Throughput Computing (CHTC) at the University of Wisconsin{\textendash}Madison,
Open Science Grid (OSG),
Extreme Science and Engineering Discovery Environment (XSEDE),
Frontera computing project at the Texas Advanced Computing Center,
U.S. Department of Energy-National Energy Research Scientific Computing Center,
Particle astrophysics research computing center at the University of Maryland,
Institute for Cyber-Enabled Research at Michigan State University,
and Astroparticle physics computational facility at Marquette University;
Belgium {\textendash} Funds for Scientific Research (FRS-FNRS and FWO),
FWO Odysseus and Big Science programmes,
and Belgian Federal Science Policy Office (Belspo);
Germany {\textendash} Bundesministerium f{\"u}r Bildung und Forschung (BMBF),
Deutsche Forschungsgemeinschaft (DFG),
Helmholtz Alliance for Astroparticle Physics (HAP),
Initiative and Networking Fund of the Helmholtz Association,
Deutsches Elektronen Synchrotron (DESY),
and High Performance Computing cluster of the RWTH Aachen;
Sweden {\textendash} Swedish Research Council,
Swedish Polar Research Secretariat,
Swedish National Infrastructure for Computing (SNIC),
and Knut and Alice Wallenberg Foundation;
Australia {\textendash} Australian Research Council;
Canada {\textendash} Natural Sciences and Engineering Research Council of Canada,
Calcul Qu{\'e}bec, Compute Ontario, Canada Foundation for Innovation, WestGrid, and Compute Canada;
Denmark {\textendash} Villum Fonden and Carlsberg Foundation;
New Zealand {\textendash} Marsden Fund;
Japan {\textendash} Japan Society for Promotion of Science (JSPS)
and Institute for Global Prominent Research (IGPR) of Chiba University;
Korea {\textendash} National Research Foundation of Korea (NRF);
Switzerland {\textendash} Swiss National Science Foundation (SNSF);
United Kingdom {\textendash} Department of Physics, University of Oxford.

\end{document}